\documentclass[12pt,preprint]{aastex}%emulateapj
\shorttitle{Metal-Poor Planet Search. I.}
\shortauthors{Sozzetti et al.}

\begin{document}


\title{A Keck/HIRES Doppler Search for Planets Orbiting Metal-Poor Dwarfs. 
I. Testing Giant Planet Formation and Migration Scenarios}


\author{Alessandro Sozzetti\altaffilmark{1,2}, 
 Guillermo Torres\altaffilmark{1}, David W. Latham\altaffilmark{1},
 Bruce W. Carney\altaffilmark{3}, Robert P. Stefanik\altaffilmark{1}, 
Alan P. Boss\altaffilmark{4}, John
B. Laird\altaffilmark{5}, and Sylvain G. Korzennik\altaffilmark{1}} 
\altaffiltext{1}{Harvard-Smithsonian Center for Astrophysics, 60
Garden Street, Cambridge, MA 02138 USA}
\altaffiltext{2}{INAF - Osservatorio Astronomico di Torino, 10025 Pino 
Torinese, Italy}
\altaffiltext{3}{Department of Physics \& Astronomy,
University of North Carolina at Chapel Hill, Chapel Hill, NC 27599 USA}
\altaffiltext{4}{Department of Terrestrial Magnetism, 
Carnegie Institution of Washington, 
5241 Broad Branch Road, NW, Washington, DC 20015 USA}
\altaffiltext{5}{Department of Physics \& Astronomy,
Bowling Green State University, Bowling Green, OH 43403 USA}
\email{asozzett@cfa.harvard.edu}
\email{gtorres@cfa.harvard.edu}
\email{dlatham@cfa.harvard.edu}
\email{bruce@physics.unc.edu}
\email{rstefanik@cfa.harvard.edu}
\email{boss@dtm.ciw.edu}
\email{laird@tycho.bgsu.edu}
\email{skorzennik@cfa.harvard.edu}


\begin{abstract}
We describe a high-precision Doppler search for giant planets 
orbiting a well-defined sample of metal-poor dwarfs in the field. 
This experiment constitutes a fundamental test of theoretical 
predictions which will help discriminate between proposed giant planet 
formation and migration models. We present here details on the survey as well 
as an overall assessment of the quality of our measurements, making use of
the results for the stars that show no significant velocity variation. 

\end{abstract}


%% subject header.  Objects should be in the appropriate "individual"
%% headers (e.g. quasars: individual, stars: individual, etc.) with the
%% additional provision that the total number of headers, including each
%% individual object, not exceed six.  The \objectname{} macro, and its
%% alias \object{}, is used to mark each object.  The macro takes the object
%% name as its primary argument.  This name will appear in the paper
%% and serve as the link's anchor in the electronic edition if the name
%% is recognized by the data centers.  The macro also takes an optional
%% argument in parentheses in cases where the data center identification
%% differs from what is to be printed in the paper.

\keywords{planetary systems: formation --- stars: statistics --- 
stars: abundances --- techniques: radial velocities}


\section{Introduction}

With a present-day catalogue of $\sim 180$ extrasolar planets
\footnote{See, for example, http://www.obspm.fr/encycl/encycl.html and 
http://www.dtm.ciw.edu/boss/iauindex.html}, 
several important statistical properties of the sample are beginning to 
emerge (for a review, see for example Marcy et al. 2005 and Udry et al. 
2006). Some of the most intriguing features unveiled so far include 
gas giant planets in few-day orbits, eccentricities spanning the entire range 
of possible values, the existence of a `Brown Dwarf Desert', correlations
among planetary orbital and physical parameters, evidence for a wide 
variety of dynamical interactions in multiple-planet systems, and a 
dependence of the frequency and properties of planetary systems on some
of the characteristics of the parent stars (mass, metallicity).

The largely unexpected properties of extrasolar planets have boosted 
theoretical research in the field of planetary system formation and 
evolution. Recently, improved understanding of some of the complex 
aspects of the underlying physical processes governing planet formation 
in gaseous protoplanetary disks has enabled us to move from attempts to 
explain already known features of the exoplanet sample to more refined 
models making testable predictions. In turn, surveys for planets with a 
variety of techniques have started targeting more carefully defined 
and selected stellar samples, with the aim of providing the observational 
evidence required to address a specific scientific question (rather than 
simply focusing on the discovery of extrasolar planets), and thus 
helping to discriminate between different theoretical planet formation models.

This is the first of a series of papers in which we present results from 
a spectroscopic search for giant planets orbiting a sample of metal-poor 
stars in the field. The two competing mechanisms for gas giant planet 
formation, core accretion and disk instability, produce significantly different 
distributions of planet masses and orbital elements (e.g., Rice et al. 2003; 
Ida \& Lin 2004a), and they predict a 
very different dependence of planetary frequency on stellar metallicity 
and mass (Boss 2002, 2006; Ida \& Lin 2004b, 2005; Laughlin et al. 2004). 
In order to determine whether core accretion or disk instability is the 
dominant formation mode for giant planets, or to verify the existence of 
bimodal planet formation, one should then compare the frequency of gas giant 
planets and their properties between metal-rich and metal-poor stars.  
However, the low-metallicity stellar sample that has been searched for 
planets is at present too small to 
test but the most outstanding differences between such hypothetical 
populations. It is now crucial to provide a statistically significant, 
unbiased sample of metal-deficient stars screened for giant planets. 

We have arranged this first paper as follows. In Section 2 we present the 
scientific case for a Doppler survey for giant planets orbiting field 
metal-poor dwarfs. We describe in Section 3 some of the technical aspects 
of our Keck/HIRES radial-velocity survey. We present in Section 4 
preliminary results from the first 3 years of observations, focusing 
on the detailed assessment of the quality of our measurements, making use of
the results for all the stars that show no significant velocity variation. 
We provide in Section 5 a brief summary and conclusions.

\section{Testing Giant Planet Formation and Migration Models}

\subsection{Stellar Metallicity and Planets: Observations}

The connection between the presence of giant planets and the metal content 
of the parent stars has been the subject of a significant number of studies 
in the past (see for example Gonzalez 2003, and references therein). 
The average of the metallicity distribution of planet-hosting stars is 
[Fe/H] $\simeq 0.14$ (e.g., Fischer \& Valenti 2005), whereas the 
mean value for the solar neighborhood is [Fe/H] $\simeq -0.1$ 
(Nordstr\"om et al. 2004). With improved statistics, in recent years 
the hypothesis that super-solar metallicity could correspond to a higher 
likelihood of a given star to harbor a planet has been conclusively 
proved (e.g., Santos et al. 2001, 2004; Fischer \& Valenti 2005).  
As of today, the frequency of giant planets around metal-rich ([Fe/H] $\gtrsim 0.3$) 
dwarfs (integrating over the F-G-K spectral types) is $f_p\gtrsim 20\%$, 
while this fraction decreases to  $f_p\simeq 3\%$ 
for metal-poor stars ($-0.5\lesssim$ [Fe/H] $\lesssim 0.0$). For metallicities 
below solar, $f_p$ appears to be roughly constant (Santos et al. 2004a; 
Fischer \& Valenti 2005). 

Many authors have debated whether the observational evidence is an indicator
of primordial high metallicity in the planet host stellar sample 
(Santos et al. 2001; Reid 2002), or if
the trend with [Fe/H] could be due to pollution by ingested
planetary material (Laughlin 2000; Gonzalez et
al.\ 2001; Israelian et al. 2001; Murray \& Chaboyer 2002). 
The idea of pollution is 
losing credit among the scientific community, primarily based on the 
lack of any correlation between [Fe/H] and effective temperature $T_\mathrm{eff}$,
or convective envelope mass $M_\mathrm{conv}$, for the planet host sample
(e.g. \citeauthor{pinsonneault01}~\citeyear{pinsonneault01}; 
\citeauthor{fischer05}~\citeyear{fischer05}; 
\citeauthor{santos03}~\citeyear{santos03},~\citeyear{santos04a}, 
and references therein). 
Results on this specific issue are not yet conclusive, however. For example,
Vauclair (~\citeyear{vauclair04}) has recently pointed out how the absence of a
[Fe/H]-$M_\mathrm{conv}$ correlation does not automatically imply that
stars with planets have not been polluted.

Furthermore, theoretical calculations
(\citeauthor{montalban02}~\citeyear{montalban02};
\citeauthor{boesgaard02}~\citeyear{boesgaard02})
suggest that detection of anomalous abundances of
rare elements such as lithium (Li) or beryllium (Be) could be
interpreted as evidence for recent accretion of planets onto the atmosphere of a star.
The abundances of Li isotopes in the spectral region around the
6707\AA\, line in planet-host stars have been investigated in the recent past
by several authors
(\citeauthor{gonzalez00}~\citeyear{gonzalez00};
\citeauthor{ryan00}~\citeyear{ryan00};
\citeauthor{isra01}~\citeyear{isra01},~\citeyear{israelian03},
~\citeyear{israelian04}; \citeauthor{reddy02}~\citeyear{reddy02};
\citeauthor{mandell04}~\citeyear{mandell04}), and similar studies have
been conducted for the \ion{Be}{2} lines at 3130\AA\, and 3131\AA.\,
(\citeauthor{garcia98}~\citeyear{garcia98};
\citeauthor{deliyannis00}~\citeyear{deliyannis00};
\citeauthor{santos02}~\citeyear{santos02},~\citeyear{santos04b}).
While the presence of the $^6$Li isotope has actually been 
detected in some planet-harboring stars 
(\citeauthor{isra01}~\citeyear{isra01},~\citeyear{israelian03};
\citeauthor{laws01}~\citeyear{laws01}), suggesting that accretion of planetary
material can indeed take place in some stars, in general stars with planets
have normal light-element abundances, typical of field stars. It thus
seems unlikely that pollution effects can be responsible for the overall
metallicity enhancement of the planet host stellar sample. 

In addition, analyses of over a dozen other elements
have been carried out in the recent past 
(\citeauthor{santos00}~\citeyear{santos00};
\citeauthor{gonzalez01}~\citeyear{gonzalez01};
\citeauthor{smith01}~\citeyear{smith01};
\citeauthor{takeda01}~\citeyear{takeda01};
\citeauthor{sadakane02}~\citeyear{sadakane02};
\citeauthor{zhao02}~\citeyear{zhao02}; 
\citeauthor{bodaghee03}~\citeyear{bodaghee03};
\citeauthor{ecuvillon04a}~\citeyear{ecuvillon04a},~\citeyear{ecuvillon04b},
~\citeyear{ecuvillon06a}; \citeauthor{beirao05}~\citeyear{beirao05}; 
\citeauthor{gilli06}~\citeyear{gilli06}; Sozzetti et al. 2006), 
and the general evidence is that the abundance 
distributions in stars with planets are the
extension of the observed behavior for [Fe/H], a result quantified by
trends of decreasing [$X$/Fe] with decreasing [Fe/H]. The absence of 
any statistically significant trend of metallicity [$X$/H] with condensation 
temperature $T_c$ (e.g., Sozzetti et al. 2006, Ecuvillon et al. 2006b; Gonzalez 2006) 
is one more piece of circumstantial evidence that 
the best explanation for the metallicity excess in stars with
planets is that the enhanced [Fe/H] is primordial in nature. 

Several studies have also focused on possible correlations
between stellar metallicity and planet properties. While no significant 
trend was found between [Fe/H] and planet mass or orbital eccentricity (e.g., 
Udry et al. 2002; Santos et al. 2001, 2003; Fischer et al. 2002), evidence 
(albeit weak) for a correlation between the metallicity of planet-harboring 
stars and the orbital periods $P$ of the planets has been pointed out (e.g., 
Sozzetti 2004, and references therein; Santos et al. 2006). This correlation 
is highlighted by an excess of close-in planets, on few-days orbits, around 
the metal-rich ([Fe/H]$\gtrsim 0.0$) sample of planet hosts. 

\subsection{Stellar Metallicity and Planets: Theory}

Within the framework of the scenario of gas giant planet formation by 
core accretion (e.g., Lissauer 1993; Pollack et al. 1996; Alibert et al. 2005), 
recent studies have successfully reproduced the strong dependence of planetary frequency 
on stellar metallicity, in qualitatively good agreement with the observed trend 
(Kornet et al. \citeyear{kornet05}; Ida \& Lin \citeyear{ida04b}). The 
probability of forming giant planets according to the disk 
instability model (e.g., Boss 1997, 2000; Mayer et al. 2004), however, 
is remarkably insensitive to the primordial surface density of solids of the
protoplanetary disk (\citeauthor{boss02}~\citeyear{boss02}; 
\citeauthor{rice03} \citeyear{rice03}),
thus planet occurrence should not be hampered around metal-poor stars. 
On the one hand, the observed trend suggests that giant planet formation 
by core accretion predominates in the metal-rich regime ([Fe/H]$\gtrsim 0.0$). 
On the other hand, $f_p$ appears to be rather flat in the metal-poor 
regime ([Fe/H]$\lesssim 0.0$). The possible evidence for bi-modality of the planet 
frequency distribution as a function of metallicity (Santos et al. 2004; Fischer 
\& Valenti 2005) suggests the existence of two different mechanisms for 
forming gas giant planets. However, due to the low numbers of metal-poor stars 
([Fe/H]$\lesssim -0.5$) surveyed to-date, no definitive conclusion can be drawn, 
except that maybe both mechanisms operate (\citealp{beer04}).

If real, several possible explanations can be put forth for the existence 
of a $P-$[Fe/H] correlation. For example, migration rates might be slowed down 
in metal-poor protoplanetary disks (\citeauthor{livio03}~\citeyear{livio03}; 
Boss 2005), 
although the predicted 
dependence of migration time-scales on [Fe/H] appears to be somewhat weak. 
The correlation may also arise as a consequence of longer
time-scales for giant planet formation around metal-poor stars, and thus
reduced migration efficiency before the disk dissipates
(\citeauthor{ida04a} \citeyear{ida04a}; Alibert et al. 2005). Another possibility 
(Santos et al. 2006) is related to planetary internal 
structure arguments: if planets formed in low-metallicity disks 
have small rocky cores (e.g., Pollack et al. 1996; Ida \& Lin 2004), 
their low density might hamper survival against 
evaporation (Baraffe et al. 2004; Lecavelier des Etangs et al. 2004), once 
they have migrated to very close-in orbits. However, even in 
this case small-number statistics in the low-metallicity regime prevents 
one from reaching a clear conclusion.

One way or another, in order to unambiguously determine 
the role of metallicity in gas giant planet formation (for example, is $f_p$ 
a truly monotonic function of [Fe/H], or is planetary frequency constant in 
the metal-poor regime?), and consequently discriminate between proposed 
explanations for the observed trends in the data, it is crucial to 
provide a statistically significant, unbiased sample of metal-poor stars 
screened for giant planets. 

\subsection{Searching for Planets Around Field Metal-Poor Stars}

The absence of short-period ($P \leq 8.3$ days) transiting planets in the moderately 
metal-deficient ([Fe/H] $\simeq -0.7$) globular cluster
47 Tucan\ae\, has been used by Gilliland et al. (\citeyear{gilliland00}) and 
by Weldrake et al. (\citeyear{weldrake05}) to argue that low-metallicity 
stars are less likely to harbor giant planets. The claims by these authors 
suffer however from some ambiguity, because 
in the cluster core investigated by Gilliland
et al. (\citeyear{gilliland00}) with $HST$ transit photometry, crowding could play a
significant role in giant planet formation, migration, and survival 
(e.g., Davies \& Sigurdsson 2001; Bonnell et al. 2001; 
Hurley \& Shara 2002; Fregeau et al. 2006).
The outer regions of the cluster monitored by Weldrake et al. (\citeyear{weldrake05})
are less affected by crowding. However, the lower occurrence rate of hot Jupiters
in a metal-poor environment does not rule out the existence of a population
of giant planets at wider radii. Indeed, this possibility has 
recently been supported by the 
findings of Sigurdsson et al. (2003), who, using HST data, were able to 
infer a mass of a few Jupiter masses for the third, long-period component 
orbiting the white dwarf - pulsar system B1620-26 in the globular cluster 
M4, five times more metal-poor than 47 Tuc. Their results provide the first 
evidence for planet formation in very metal-poor environments. 
In light of the M4 announcement, the 47 Tuc results could be 
re-interpreted as follows: Dynamical disruption in dense clusters is not 
sufficient to completely destroy any planetary population, and the lack 
of transiting planets on short-period orbits might be due to other processes, 
such as a metallicity dependence in the migration mechanism, or 
suppression of migration (but not formation) in globular clusters
\footnote{Notably, the lack of any apparent correlation in the plane 
defined by minimum mass $M_p\sin i$ and orbital period $P$ for planets in binary 
and triple stellar systems (as opposed to the observed $M_p\sin i-P$ 
correlation in the case of planets orbiting single stars) is interpreted 
as evidence for enhanced migration efficiency for planets formed in 
stellar systems (Zucker \& Mazeh 2002; Udry et al. 2003; Eggenberger 
et al. 2004). However, the birthplaces for such systems are 
stellar groups and clusters with much lower stellar densities than 
globular clusters. Thus, both the impact on planet formation and 
migration efficiency as well as the evolutionary history of dynamical 
interactions are likely not the same in these two different environments 
(e.g., Davies \& Sigurdsson 2001; Adams et al. 2006)}. 

By addressing the field population of metal-poor stars directly, it is 
then possible to eliminate dynamical interactions in dense stellar 
environments as a possible source of interference with planet formation, 
or with migration to close-in orbits, or with planet survival.  

\section{The Keck/HIRES Doppler Survey of Metal-Poor Dwarfs}

In 2003, we began a high-precision radial-velocity survey of 
$\sim 200$ metal-poor stars, using HIRES on the Keck 1 telescope 
(Vogt et al. 1994) and its I$_2$ gas absorption cell as the reference 
velocity metric (Butler et al. 1996). The goal of this project is to carry out an initial 
reconnaissance for gas giant planets orbiting within 1 AU of a statistically 
significant sample of low-metallicity dwarfs. 

The sample has been drawn from 
the Carney-Latham and Ryan samples of metal-poor, high-velocity field stars 
(e.g., Carney et al. 1994; Ryan 1989; Ryan \& Norris 1991). 
The initial combined database totaled 
1558 objects. A number of selection criteria have been adopted in order to 
finalize our list of targets. First of all, a key advantage of the Carney-Latham and 
Ryan samples is that we have monitored the radial velocities of the stars 
in these samples, mostly for more than 3,000 days, using the CfA Digital Speedometers 
(Latham 1992). This has allowed us to identify most of the stars with stellar
companions that would interfere with the formation or survival of 
planets in the habitable zones (Carney et al. 2001; Latham et al. 2002). 
All stars included in the final list of targets showed no sign of velocity 
variation at the 0.5 to 1.0 km s$^{-1}$ level. Second, significant chromospheric 
activity (quantified for example through the chromospheric emission ratio 
$R^\prime_\mathrm{HK}$) and large values of stellar rotational velocity $v\sin i$ 
should be avoided, as they constitute sources of intrinsic radial velocity ``jitter'' 
that can mask, and sometimes even mimic, the presence of orbital reflex motion 
due to planetary mass companions (Saar et al. 1998; Santos et al. 2000; 
Queloz et al. 2001; Paulson et al. 2004). 
Fortunately, old stars have the advantage that they rotate slowly and have low
levels of chromospheric activity.  All of the stars in our sample
exhibit rotational velocities $v\sin i\leq 10$ km s$^{-1}$, and most of them
have rotations below the resolution limit of the CfA Digital Speedometers 
(8.5 km s$^{-1}$), so that a value of $v\sin i$ could not be
determined. Thus, we do not expect velocity jitter due to astrophysical 
phenomena to be a problem for this sample. 

However, metal-poor stars have weak absorption lines in comparison to 
their solar-metallicity counterparts. The lines also grow weaker as the
effective temperature rises. Furthermore, very metal-poor stars are
rare, and therefore they tend to be distant and faint. In order to 
characterize the behavior of the radial velocity precision $\sigma_\mathrm{RV}$ 
as a function of stellar metallicity, effective temperature $T_\mathrm{eff}$, 
and visual magnitude $V$ (assuming non-rotating, inactive stars), we have
run simulations utilizing the CfA library of synthetic stellar spectra 
(e.g., Nordstr\"om et al. 1994; Latham et al. 2002). 
We show in Figure~\ref{rvdegr} how $\sigma_\mathrm{RV}$ degrades as a 
function of [Fe/H] and $T_\mathrm{eff}$, assuming a fixed exposure time 
and a typical measurement precision of 10 m s$^{-1}$ for [Fe/H] $= 0.0$. 
For a solar-type star, [Fe/H] $=-1.0$ corresponds to a degradation in 
$\sigma_\mathrm{RV}$ of a factor $\sim 2$, while for a significantly cooler 
star, with a more complex spectrum\footnote{The velocity information content 
depends on the mean absolute value of the slope of the spectrum, which 
increases for later spectral types (e.g., Butler et al. 1996; 
Bouchy et al. 2001)}, the effect is less severe (in 
addition, surface gravity constitutes only a higher-order effect). This result 
confirms the empirical findings of Santos et al. (2003) and Fischer \& Valenti 
(2005), based on the analyses of the achieved velocity precision as a function of 
metallicity carried out with the stellar databases of 
their respective Doppler planet surveys. Those studies were aimed at 
ruling out possible observational biases that might contribute to 
the observed correlation between $f_p$ and [Fe/H]. In 
both cases, Santos et al. (2003) and Fischer \& Valenti (2005) conclude that, 
given the typical single-measurement precision $\sigma_{\mathrm{RV}}\simeq 3-5$ m s$^{-1}$ 
achieved on bright objects (typically $V\lesssim 8.0$), a velocity degradation of a factor 
$1.5-2$ does not imply that a fraction of the giant planets orbiting low-metallicity 
stars might have gone undetected (at least in the orbital period range 
presently covered by Doppler surveys). However, not many bright metal-poor 
stars ($-0.5 \lesssim$ [Fe/H] $\lesssim 0.0$) are found in the solar 
neighborhood ($d\lesssim 50$ pc). In order to create 
a statistically significant database (hundreds of stars) of metal-poor 
([Fe/H] $\lesssim -0.5$) dwarfs, 
one must then include more distant, and fainter, objects. For the 
purpose of optimizing the trade-off between number of objects surveyed 
and total observing time required, we have refined our sample of metal-poor 
dwarfs from the Carney-Latham and Ryan surveys by
adopting magnitude and temperature cut-offs ($V\leq 12.0$ and 
$T_\mathrm{eff}\leq 6000$ K, respectively), and by selecting objects in the 
metallicity range $-2.0 \leq$ [Fe/H] $\leq -0.6$. 

In the empirical error model we have obtained using the simulations 
with the CfA database of synthetic spectra, the radial-velocity uncertainty 
is 
\begin{equation}
\sigma_\mathrm{RV}=\left(\frac{t_0}{t_\mathrm{exp}}\, 
10.^{(V-V_0)/2.5)}\right)^{1/2}\,F(\mathrm{[Fe/H]},T_\mathrm{eff}),
\end{equation}
where $F(\mathrm{[Fe/H]},T_\mathrm{eff})$ is an empirical function of 
temperature and metallicity based on the simulation results shown 
in Figure~\ref{rvdegr}, 
and $t_0$ and $V_0$ are reference integration time and magnitude 
for a star with the temperature and metallicity of the Sun. 
Based on our experience with
solar neighborhood G dwarfs observed with HIRES for the G Dwarf Planet
Search Program (Latham 2000), we have set a threshold of 20 m s$^{-1}$
precision for planet detection, and computed the relative
exposure times needed to achieve such precision, for each star in our
sample. Furthermore, we decided to limit the maximum exposure times 
to 15 minutes, to minimize uncertainties in the barycentric correction. 
We show in Figure~\ref{targdistr} the distribution of [Fe/H], $V$, and 
distance estimates (for those objects with Hipparcos 
parallaxes, photometric otherwise) 
for the sample of 278 metal-poor stars derived after 
adopting all the selection criteria described above. In order to 
highlight the different ranges of metallicity, magnitude, and distance 
spanned by our survey, the same distributions for a large fraction 
of the present-day sample of planet hosts are also shown.  
Given that the average stellar mass in our sample is $\sim 0.69$
$M_\odot$, setting a velocity precision threshold at 20 m s$^{-1}$ is
sufficient to detect (at the 5-$\sigma$ level) velocity variations of
planetary companions with minimum mass in the average
range $0.59\, M_\mathrm{J}\leq M_\mathrm{p}\sin i\leq 2.75\, M_\mathrm{J}$, 
or higher, for orbital periods in the range $0.01\leq P\leq 1$ yr. 

Finally, if it turns out to be true that planets did not form around metal-poor
dwarfs, then we need to observe a large-enough sample so that a null
result, i.e. no detections, is significant. The rate at which giant
planets with orbits inside 1 AU are being discovered by radial 
velocities appears to be about $P\simeq 3\%$ (e.g., Marcy et al. 2005). In
order for the failure to detect any planetary companions to be
significant at the 3-$\sigma$ level (corresponding to a probability of
0.0027), we need to survey a sample of $N$ stars, where
$(1-P)^N = 0.0027$, which is satisfied for $N = 194$. We therefore 
defined our final target list by selecting a subset of 200 
metal-poor stars out of the abovementioned larger sample of 278 objects, 
which will eventually provide a robust 3-$\sigma$ null result in the 
case of no detections.

\section{Preliminary Results}

Our observing program has been awarded an average of 2 
Keck/HIRES nights per semester, starting in early 2003. 
Good temporal coverage is thus a serious issue, as such 
a scheduling of the observations is good for sampling 
long-period radial-velocity variations, but rather poor 
for identifying possible short-period variables. We have 
tried to obtain at least three velocity measurements 
per star per year, with at least one set of back-to-back 
observations (taken in two consecutive nights), 
in order to mitigate our bias toward 
poor sampling (and thus significant aliasing and 
ambiguities) at short orbital periods.

The first important step is to provide an assessment of 
the long-term stability of the velocity zero-point 
and single-measurement precision achieved for planet detection, 
in light of the predicted exposure times needed to 
reach $\sigma_\mathrm{RV} = 20$ m s$^{-1}$. 
Our analysis pipeline incorporates the full 
modeling of temporal and spatial variations of the 
HIRES instrumental profile (Valenti et al. 1995), 
similarly to the method adopted by other groups (e.g., Butler et al. 1996; 
Korzennik et al. 2000; Cochran et al. 2002). Spectral modeling of each 
echelle order containing I$_2$ lines is carried out 
independently, and internal uncertainties for each observation 
are computed from the scatter of the velocities around the mean. 
This analysis technique 
has allowed us to significantly improve upon our initial estimates of 
achievable radial velocity precision. 
In Figure~\ref{rmsdistr} we show the histogram of the rms velocity 
residuals for about 80\% of our sample, for which 
we have achieved substantially uniform temporal coverage 
(typically 5-6 observations per star, with at least two 
back-to-back exposures, over a time-span of at least a year). 
The rms velocity residuals distribution 
of the {\it full} sample (excluding variables with rms $\ge 30$ m s$^{-1}$) 
averages $\sim 9$ m s$^{-1}$. In Figure~\ref{rms_vs_span} we show the 
rms velocity residuals as a function of the time-span of the 
observations. Overplotted are the 
median and standard deviation in 500-days bins (again, variable 
stars are not included). As about two dozen of the stars in our sample 
are in common with the G dwarf planet survey of Latham (2000), we could 
establish the long-term stability of the velocity zero-point over 
time-scales of up to eight years. This demonstrates the true 
radial-velocity precision we are obtaining on the sample of metal-poor 
stars, with a significant improvement of over a factor of 2 with respect 
to the targeted 20 m s$^{-1}$ single-measurement precision. 

The exposure times predicted by the model derived from the simulations 
with the CfA library of stellar spectra are determined as a 
function of [Fe/H], $T_\mathrm{eff}$, and $V$. However, our 
program stars are up to 3-4 magnitudes fainter and up to over 
100 times more metal-poor than typical targets of Doppler 
planet searches. One possible matter 
of concern would then be the evidence of systematic trends in the 
velocity scatter as a function of these three parameters. 
However, as shown in Figure~\ref{trends}, no clear rms velocity trends as a 
function of [Fe/H], $T_\mathrm{eff}$, and $V$ are present. This 
gives us confidence that the model we developed for the 
dependence of the radial velocity precision on the above 
parameters is robust. 

A more quantitative, and challenging, test to demonstrate that 
the character of our errors is well understood can be carried out 
by studying the distribution of the velocity residuals compared 
with their formal uncertainties. We define the ratio 
$\Delta\mathrm{RV}/\sigma_\mathrm{RV}$ as the difference between 
the velocity values and their mean value for each star, divided 
by their estimated uncertainties. In the ideal case, 
if internal errors are an accurate tracer of the actual uncertainties 
in the measurements, 
this ratio should have a Gaussian distribution with zero mean and unit 
dispersion. Discrepancies between predicted and actual errors 
should be reflected in measurable departures from gaussianity. 
We show in Figure~\ref{rv_sigv} the histogram of the ratio 
$\Delta\mathrm{RV}/\sigma_\mathrm{RV}$ for all our program stars 
(including variables). 
Overplotted is a reference Gaussian with zero mean and unit dispersion. 
The $\Delta\mathrm{RV}/\sigma_\mathrm{RV}$ distribution is very close 
to Gaussian, with no apparent positive or negative biases. Slightly 
elevated wings ($\sim 6\%$ of the velocity differences are larger 
than 5 sigma) indicate the presence of either non-Gaussian outliers 
or true variables. 
The result shown in Figure~\ref{rv_sigv} demonstrates that our internal 
errors are realistic. The fundamental conclusion is that we 
achieve sufficient radial-velocity precision in our sample 
to clearly detect Jupiter-mass objects within 1 AU of 
metal-deficient dwarfs. 

We show in Figure~\ref{velres} some examples of our velocity 
measurements of metal-deficient dwarfs. The top two panels 
show results for two of the stars (HD 157089 and G68-30) 
in common between our survey and the G Dwarf Planet Search program 
(Latham 2000). These objects are constant to $\sim 10$ m s$^{-1}$ 
over a time-scale of $\sim 8$ years. 
The examples of velocity time-series shown in the other 
panels of Figure~\ref{velres} cover a range of $\sim 1000$ K 
in $T_\mathrm{eff}$, $\sim 4$ magnitudes, and $\simeq 1$ dex 
in [Fe/H]. The rms of the observations range between 
6 m s$^{-1}$ and 11 m s$^{-1}$, with an average of 
$\sim 9$ m s$^{-1}$. While a large fraction of our sample 
shows no significant velocity variations over the time-span 
of the observations, a number of objects do exhibit velocity 
variability indicative of the presence of companions. 
A thorough analysis of our planet detectability thresholds 
and a detailed presentation of all our velocity measurements 
will be the subject of our second paper.

\section{Conclusions}

One way to distinguish between proposed models of gas giant 
planet formation is to confirm or rule out, on an observational 
basis, their different predictions on planet frequency $f_p$ 
as a function of the metallicity of the stellar hosts. To address 
this issue, we have undertaken a Doppler search for giant planets 
within 1 AU of a sample of 200 metal-deficient ($-2.0 \leq$ [Fe/H] $\leq -0.6$) 
dwarfs in the field. This is the sample size needed to provide 
a statistically significant result (at the $3-\sigma$ level) in the 
case of no detections. Using the Keck 1 telescope and its HIRES spectrograph, 
we have achieved a long-term radial-velocity precision of $\sim 9$ m s$^{-1}$, 
independently of [Fe/H] and $T_\mathrm{eff}$, and for stars 2 to 4 
magnitudes fainter than the targets for most other radial-velocity 
planet surveys. We have provided convincing evidence that our internal 
errors estimates are reliable, and thus demonstrated that we 
achieve sufficient radial-velocity precision in our sample 
to clearly detect Jupiter-mass objects within 1 AU of 
metal-deficient dwarfs. A number of objects with significant radial-velocity 
trends have been identified, for which we plan to perform 
follow-up observations. At the conclusion of our Doppler survey, 
we will be able to place useful upper limits on the existence of 
planetary companions of given mass and period around metal-poor stars, 
and we will then compare the frequency of gas giant planets and their 
properties between metal-rich and metal-poor stars. 
These issues will be addressed in future papers. 

\acknowledgments

A.S. acknowledges support from the Keck PI Data Analysis Fund (JPL 1267110). 
G.T. acknowledges partial support for this work from NASA Origins grant 
NNG04LG89G. J.L. is partially supported by an NSF grant AST-0307340. 
B.C. gratefully acknowledges support from an NSF grant AST-0305431.
It is a pleasure to acknowledge Mike Kurtz for very stimulating discussions.  
Special thanks are due to Dimitar Sasselov for lending initial impetus and 
support to this investigation. 
The data presented herein were obtained at the W.M. Keck Observatory, which
is operated as a scientific partnership among the California
Institute of Technology, the University of California and the
National Aeronautics and Space Administration. The Observatory was
made possible by the generous financial support of the W.M. Keck
Foundation. The authors wish to recognize and acknowledge the very
significant cultural role and reverence that the summit of Mauna
Kea has always had within the indigenous Hawaiian community. 
Without their generous hospitality, the Keck observations presented 
here would not have been possible.

\clearpage

\figcaption{Degradation in the radial velocity 
precision $\sigma_\mathrm{RV}$ as a function of stellar metallicity, 
effective temperature, and gravity, for fixed exposure 
time. The zero point of $\sigma_\mathrm{RV}$ for solar 
values of [Fe/H], $T_\mathrm{eff}$, and $\log g$ is arbitrarily 
scaled to 10 m s$^{-1}$.\label{rvdegr}}

\figcaption{Distributions of visual magnitudes (top), distances 
from the Sun (center), and metallicities (bottom) for a 
sample of 278 metal-poor stars selected with the criteria 
detailed in the text. For comparison, the same distributions 
for a large sample of 119 planet hosts are also shown 
(data from Santos et al. 2004a, 2005, and Sozzetti et al. 2004).
\label{targdistr}}

\figcaption{Rms velocity distribution for $\sim 80\%$ of 
the stars in our sample, for which uniform temporal 
coverage has been obtained (see text for details). 
Objects exhibiting significant radial velocity variations 
($> 30$ m s$^{-1}$) are not shown.\label{rmsdistr}}

\figcaption{Rms velocity residuals as a function of the time-span 
of the observations. Overplotted are the median (large filled circles) 
and standard deviation in 500-days bins. Variables with rms $> 30$ m s$^{-1}$ 
are not taken into account.
\label{rms_vs_span}}

\figcaption{Radial velocity scatter (excluding variables with 
rms $\geq 30$ m s$^{-1}$) 
as a function of [Fe/H] (left), $T_\mathrm{eff}$ (center), and $V$ (right).
Based on a rank-correlation test, the probability of no correlation 
in the three cases is 0.61, 0.14, and 0.09, respectively (i.e., no 
significant correlation is present).
\label{trends}}

\figcaption{Histogram of all the velocity residuals (including 
variables) normalized by 
their formal uncertainties. The dotted line represents a reference 
Gaussian distribution with zero mean and unit dispersion. If, for 
example, internal errors are over- or under-estimated, this effect 
should show up as distortions in the distribution. The width 
of the distribution is very close to unity, 
indicating the absence of significant biases. 
Formal errors are a good estimate of the true underlying uncertainties.
\label{rv_sigv}}

\figcaption{Observed relative velocities $\delta$RV 
for a sample of stars in our program. 
The top two panels show objects with constant 
velocity to $\sim 10$ m s$^{-1}$ over a time-span of 8 years, 
after combining observations taken in the context of the 
G Dwarf Planet Search program (Latham 2000). The mean rms 
of the stars shown in the other panels is 9 m s$^{-1}$, 
the same as the average of the full sample (excluding variable 
stars). Comparable precision is achieved over ranges of 1000 K 
in $T_\mathrm{eff}$, 4 magnitudes, and 1 dex in metallicity.
\label{velres}}

\clearpage

%% This example uses \plotone to include an EPS file scaled to
%% 80% of its natural size with \epsscale. Its caption
%% has been written to indicate that additional figure parts will be
%% available in the electronic journal.

\begin{figure}
\plotone{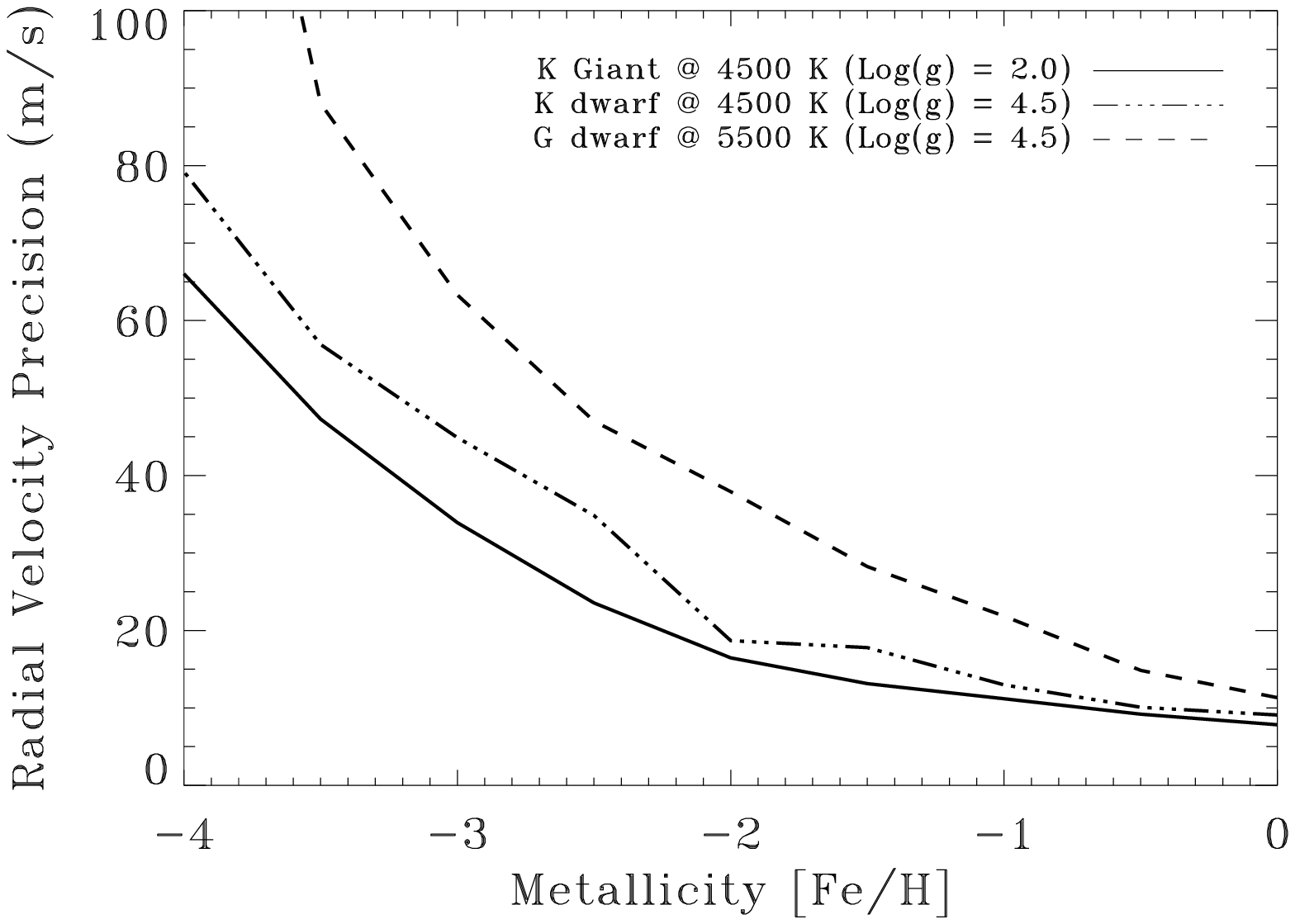}
\end{figure}

\clearpage

\begin{figure}
\centering
$\begin{array}{c}
\includegraphics[width=0.5\textwidth]{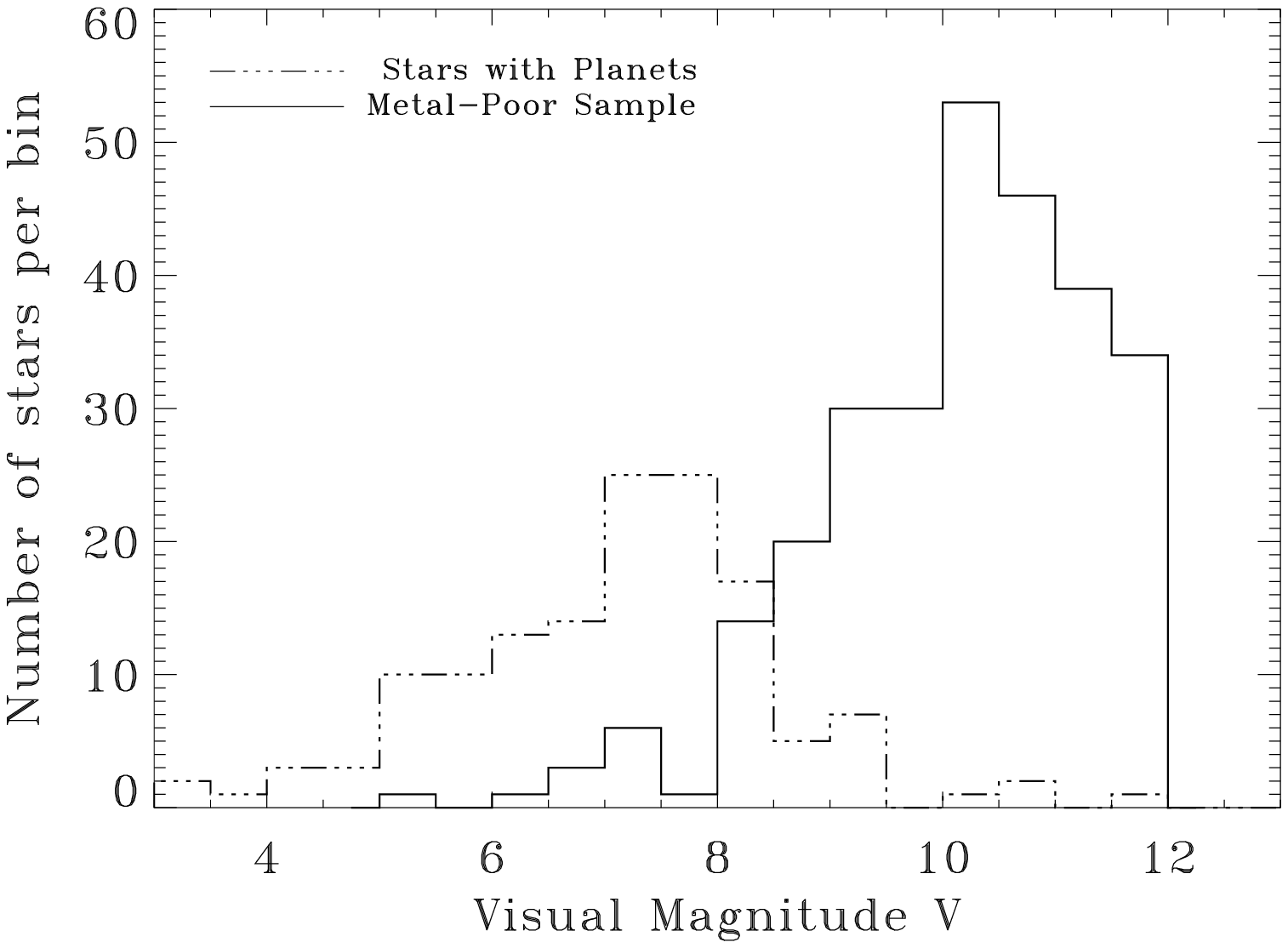} \\ 
\includegraphics[width=0.5\textwidth]{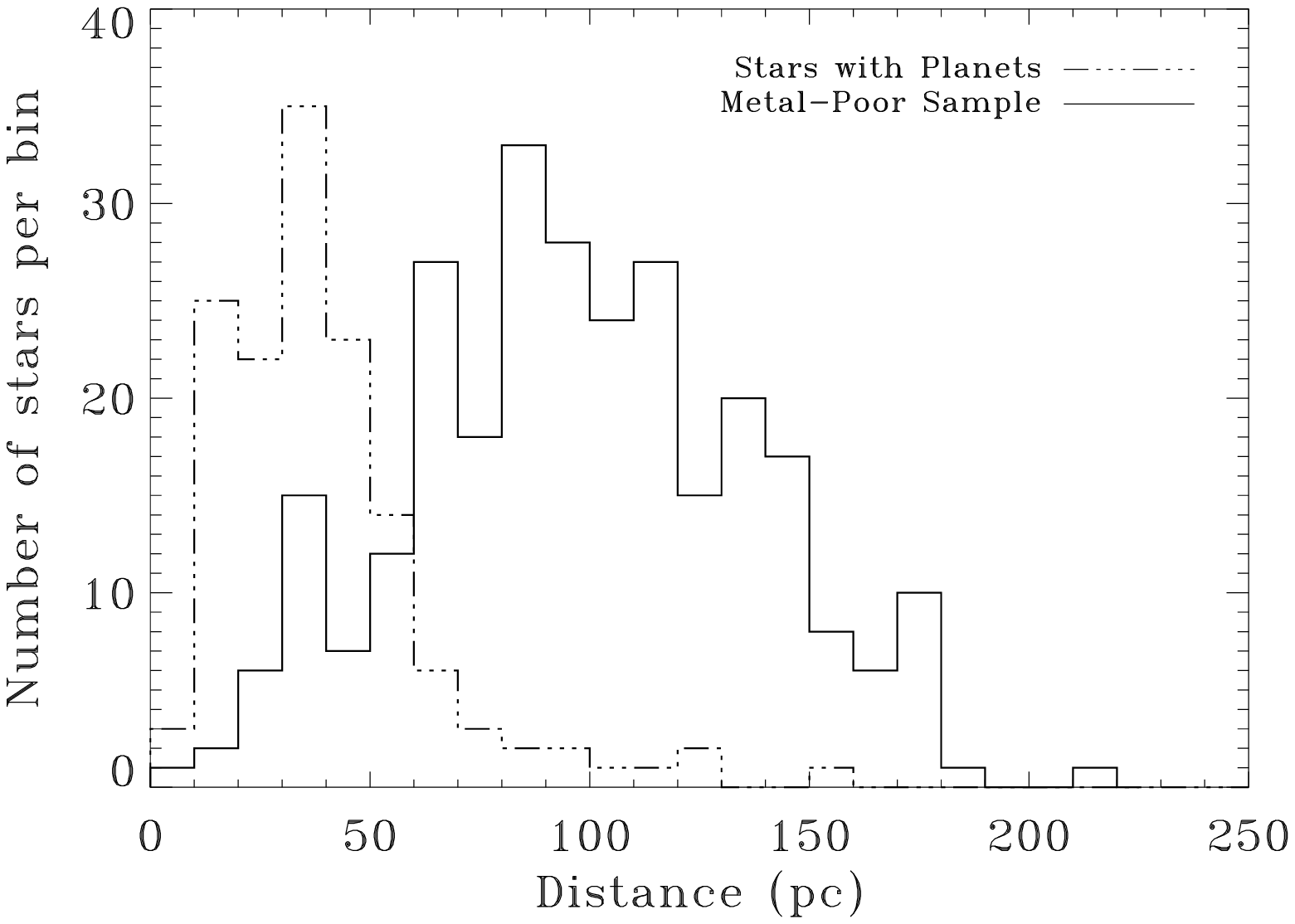} \\
\includegraphics[width=0.5\textwidth]{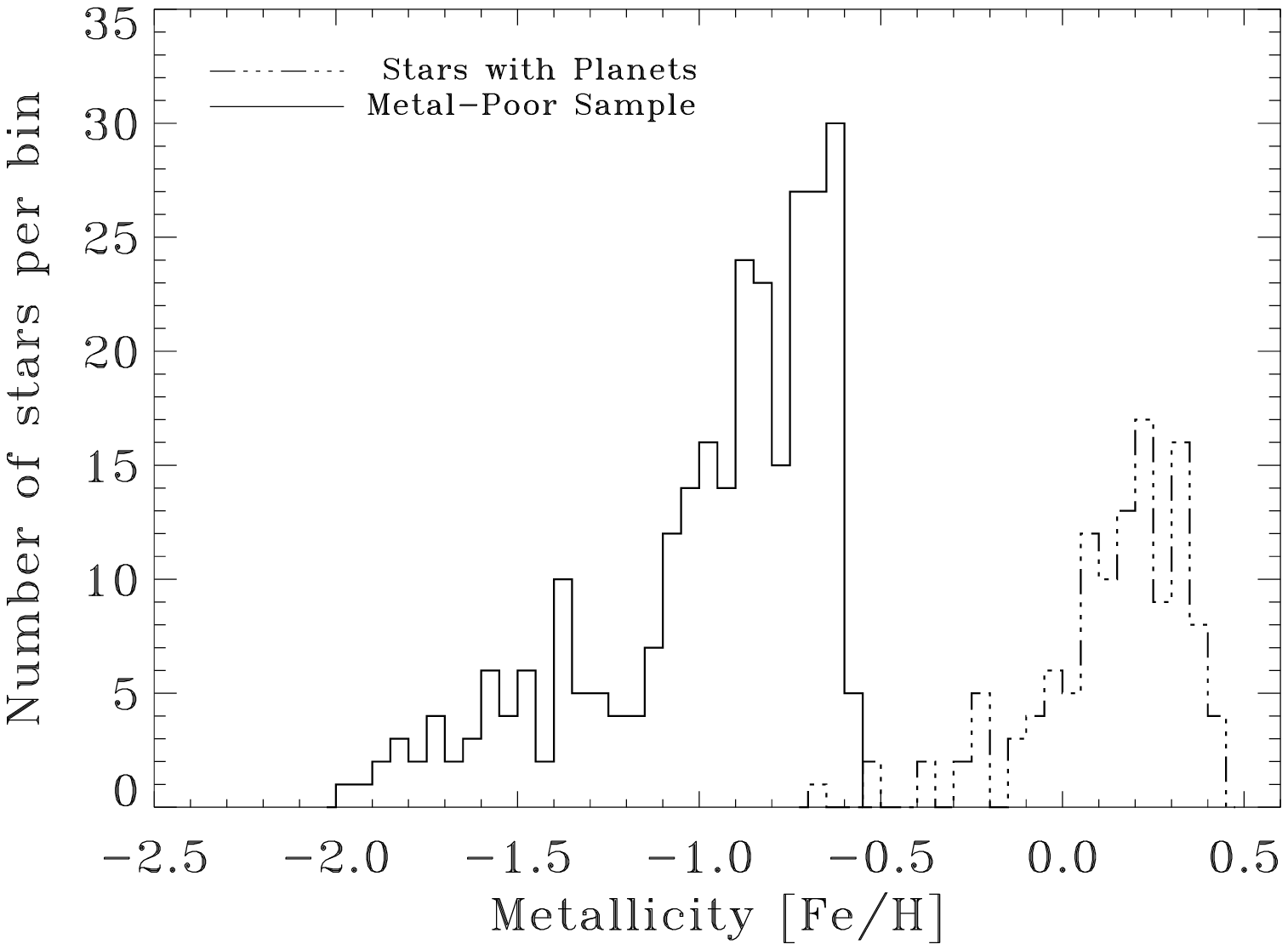} 
\end{array} $
\end{figure}

\clearpage

\begin{figure}
\plotone{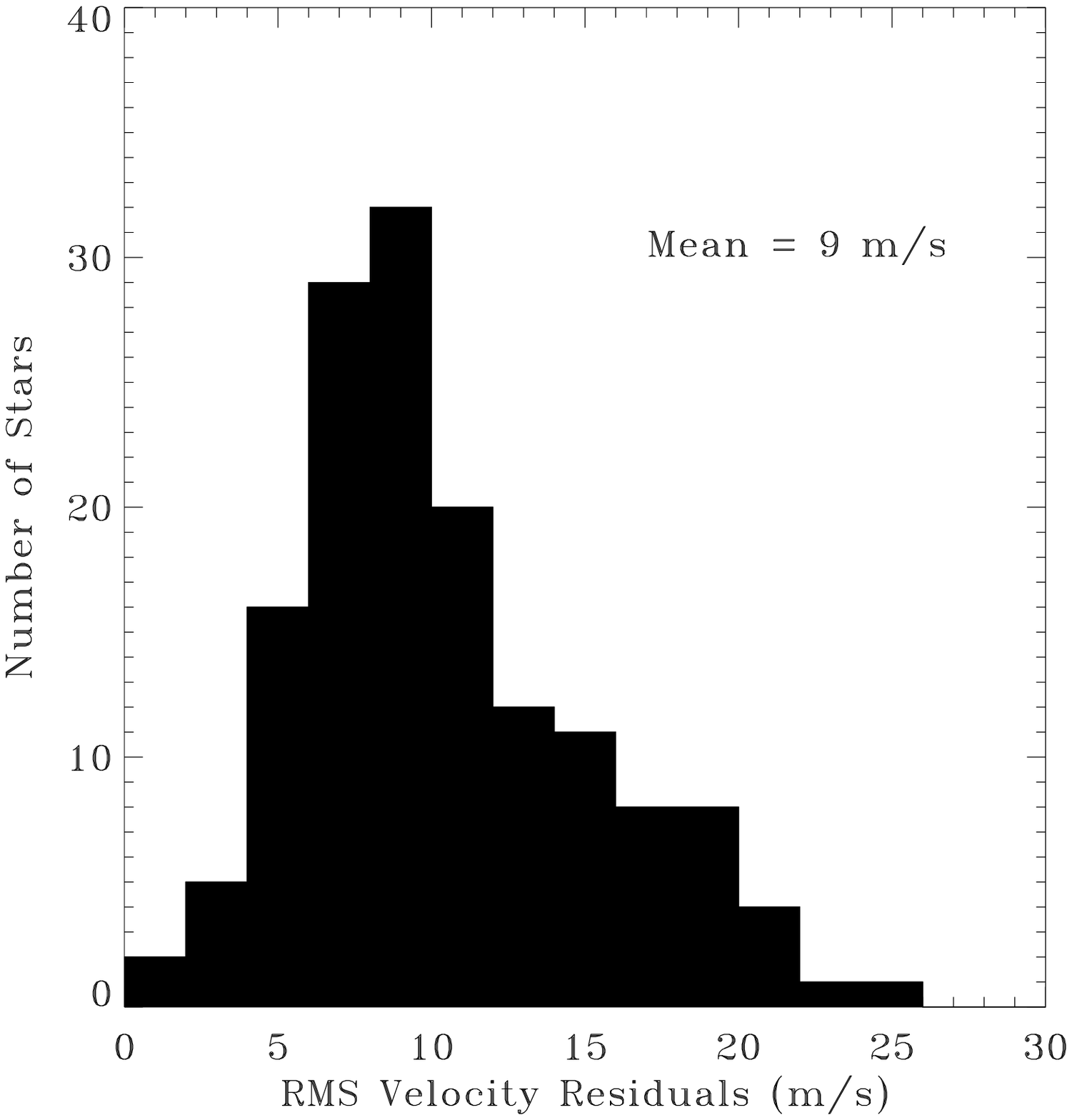}
\end{figure}

\clearpage

\begin{figure}
\plotone{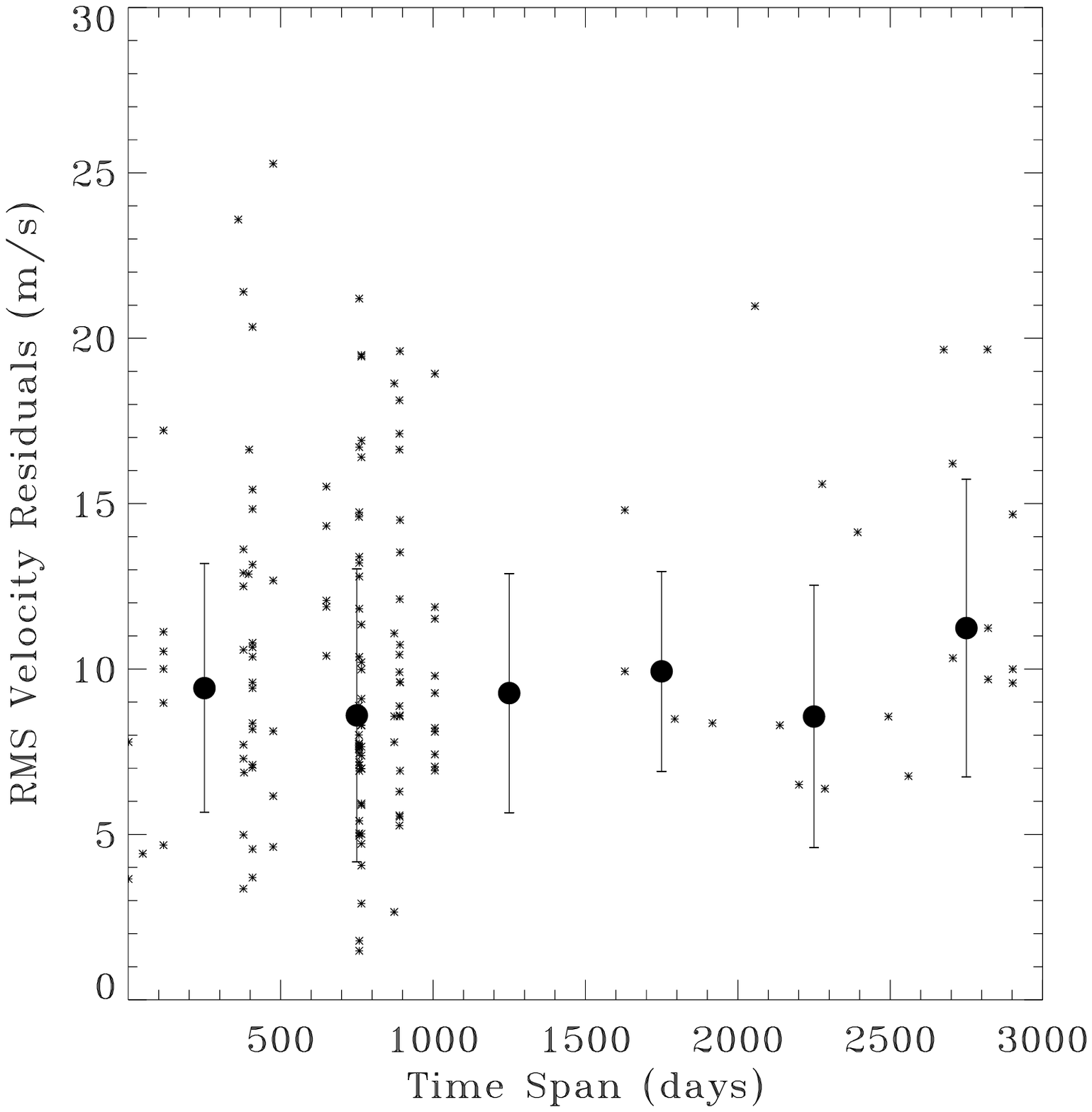}
\end{figure}

\clearpage

\begin{figure}
\centering
$\begin{array}{ccc}
\includegraphics[width=0.32\textwidth]{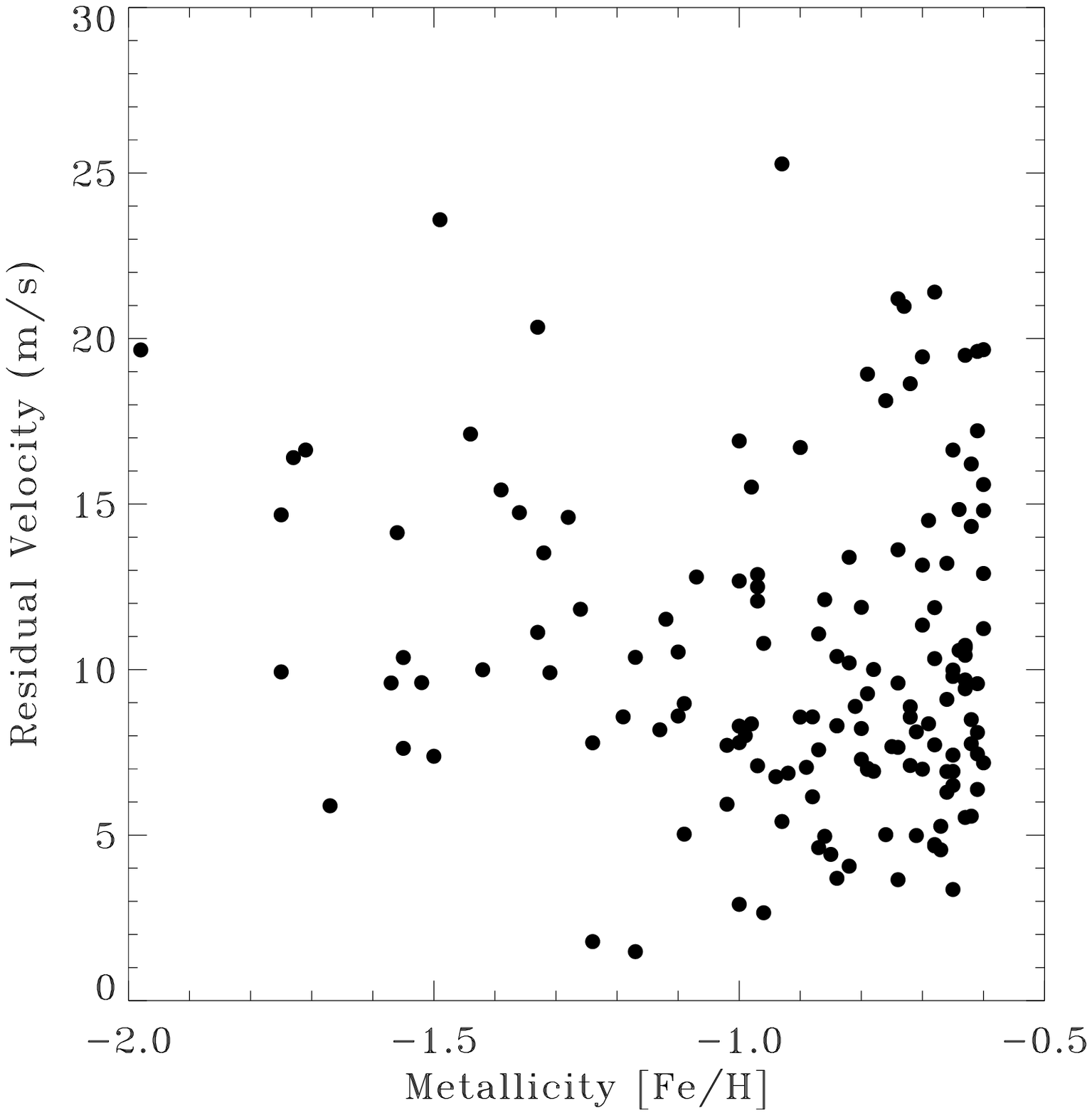} & 
\includegraphics[width=0.32\textwidth]{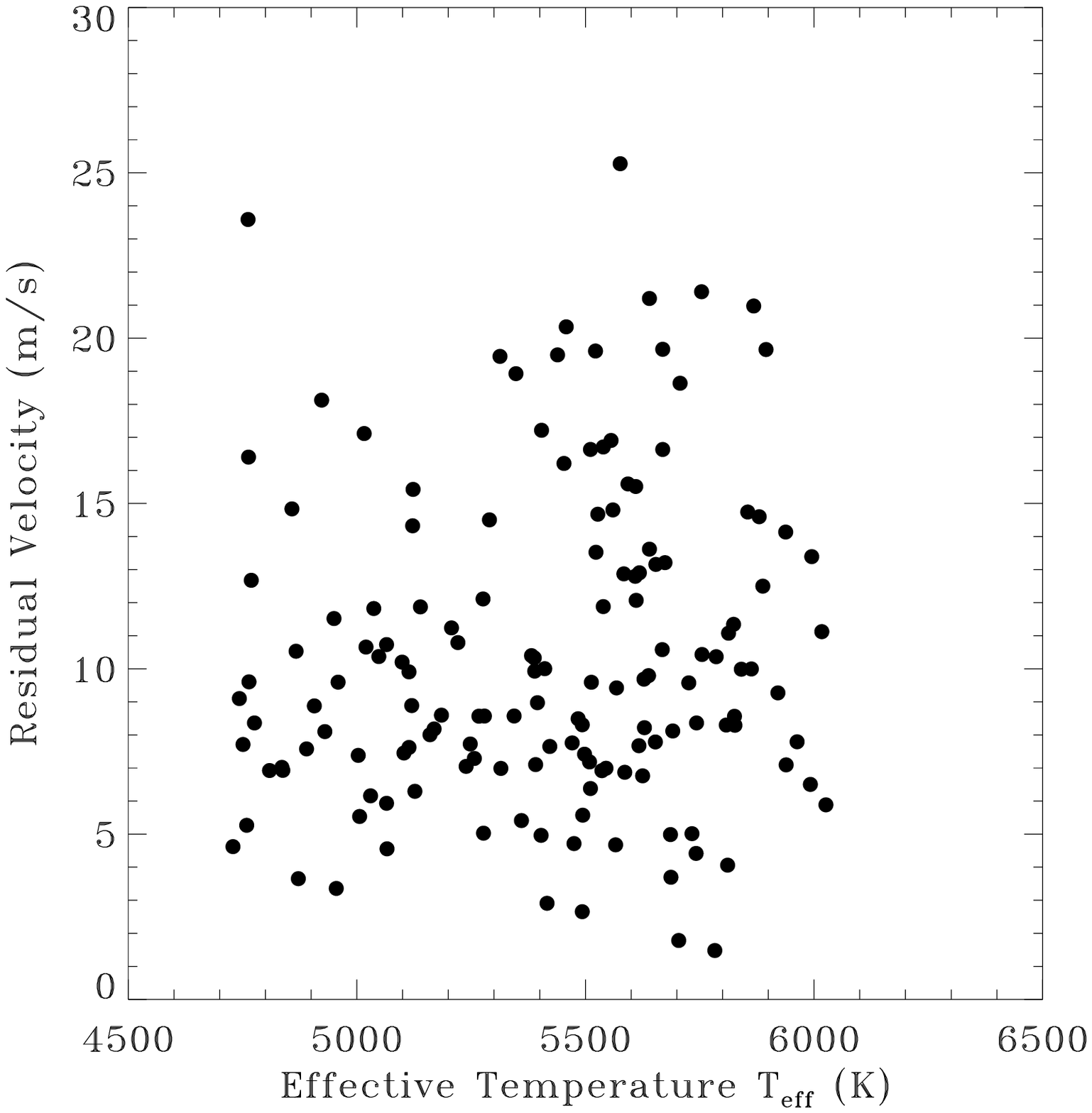} & 
\includegraphics[width=0.32\textwidth]{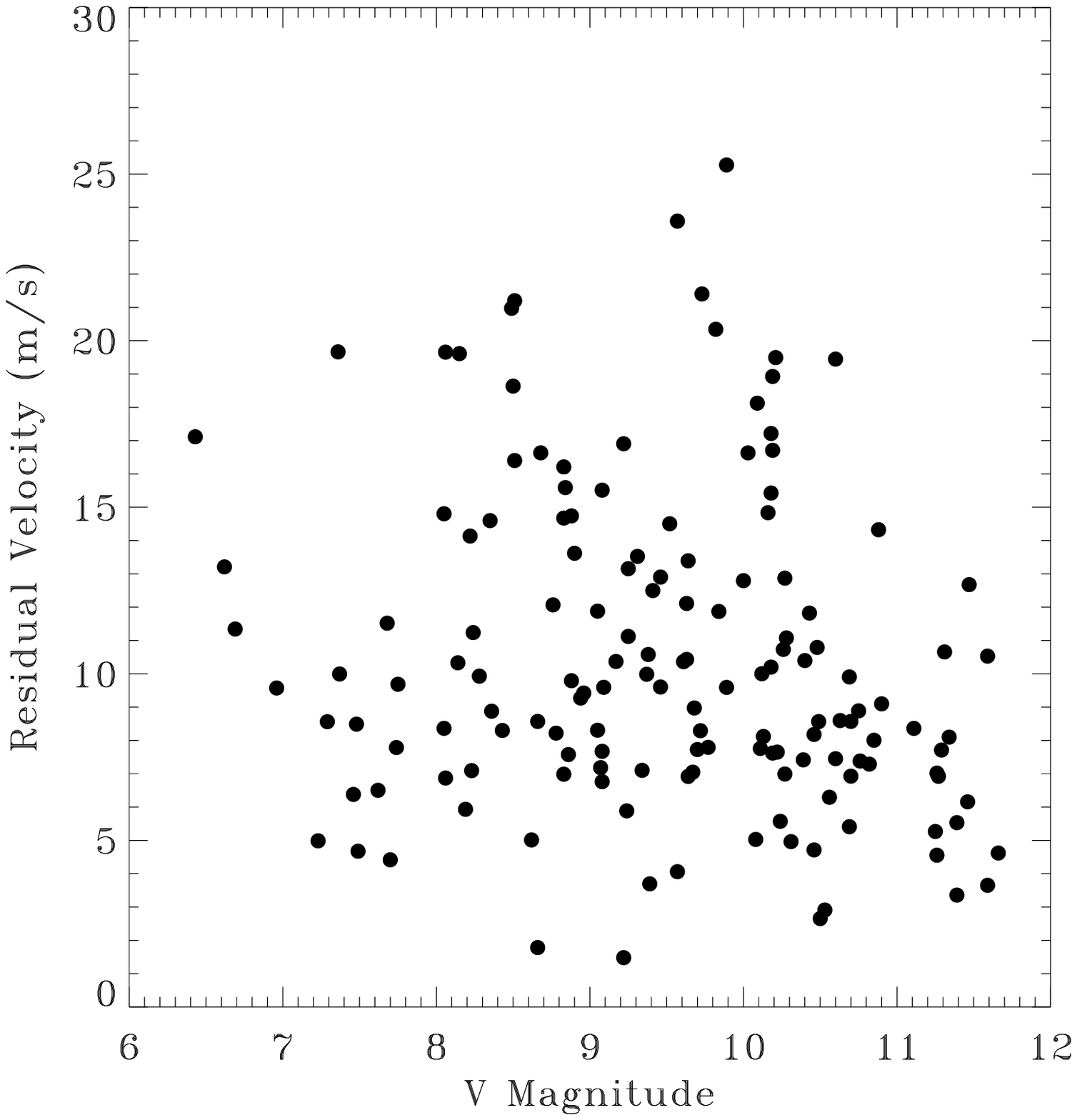} 
\end{array} $
\end{figure}

\clearpage

\begin{figure}
\plotone{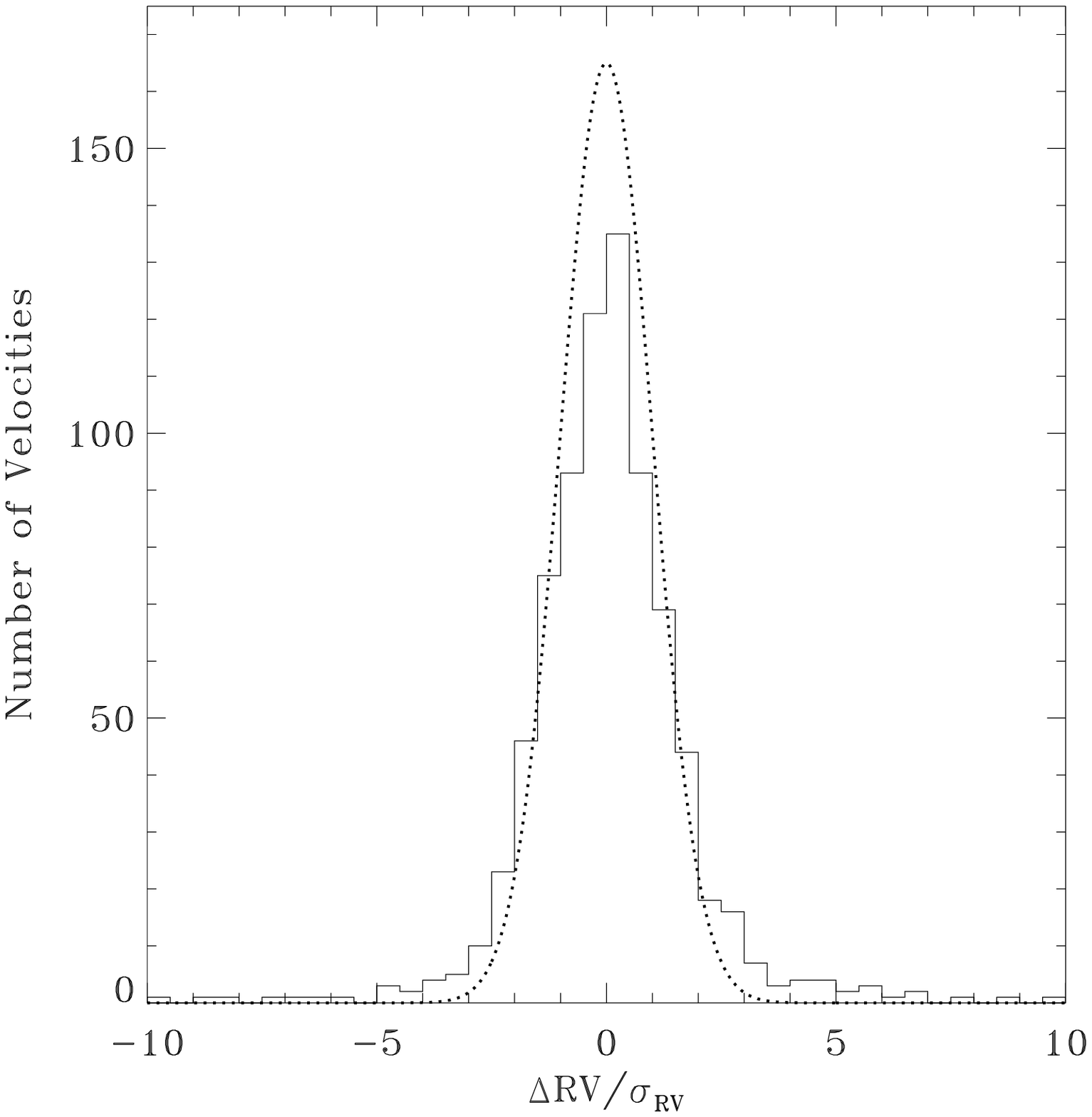}
\end{figure}

\clearpage

\begin{figure}
\plotone{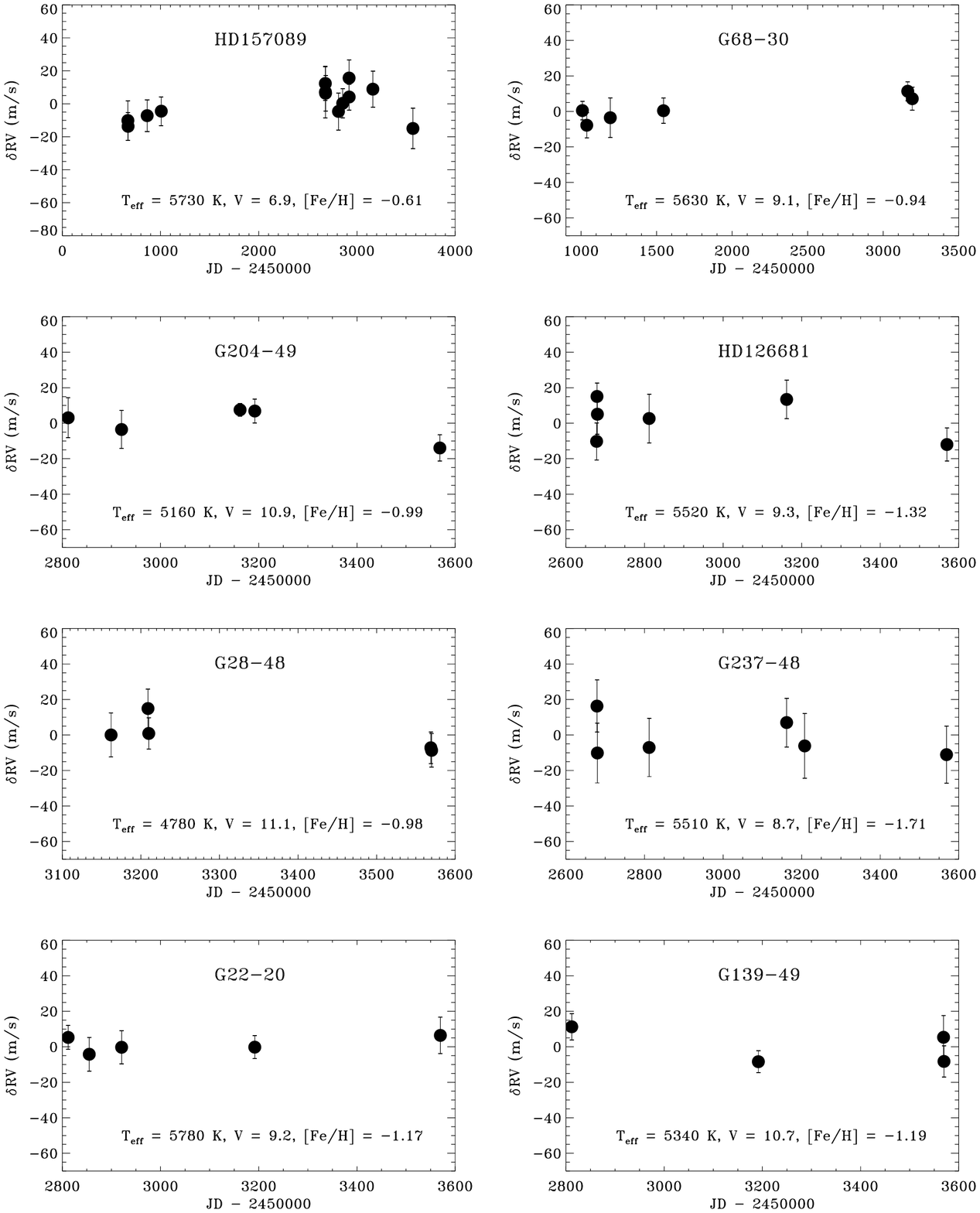}
\end{figure}

\end{document}